\documentclass[aps,pre,reprint,groupedaddress]{revtex4-1}

\usepackage{graphicx}
\usepackage{amsmath}
\usepackage{amsfonts}

\begin{document}

\title{Jet direction in bubble collapse within rectangular and triangular channels}

\author{Lebo Molefe$^{1,2}$}
\author{Ivo R. Peters$^1$}
\email[]{i.r.peters@soton.ac.uk}
\affiliation{$^1$Faculty of Engineering and Physical Sciences, University of Southampton, Southampton SO17 1BJ, UK}
\affiliation{$^2$The University of Chicago, Chicago, Illinois 60637, USA}

\date{\today}

\begin{abstract}
A vapor bubble collapsing near a solid boundary in a liquid produces a liquid jet that points toward the boundary. The direction of this jet has been studied for boundaries such as flat planes and parallel walls enclosing a channel. Extending these investigations to enclosed polygonal boundaries, we experimentally measure jet direction for collapsing bubbles inside a square and an equilateral triangular channel. Following the method of Tagawa and Peters (Phys. Rev. Fluids \textbf{3}, 081601, 2018) for predicting the jet direction in corners, we model the bubble as a sink in a potential flow and demonstrate by experiment that analytical solutions accurately predict jet direction within an equilateral triangle and square. We further use the method to develop predictions for several other polygons, specifically, a rectangle, an isosceles right triangle, and a $30^{\circ}$-$60^{\circ}$-$90^{\circ}$ right triangle.
\end{abstract}

\maketitle

\section{Introduction}
Cavitation is the sudden formation of bubbles in liquid that occurs when absolute pressure in the liquid decreases below its vapor pressure. Due to surface tension, absolute pressure must actually drop to a threshold far below the vapor pressure, to large negative pressures, for cavitation to occur. The negative threshold pressure observed in experiment varies depending on the amount of impurities in the liquid \cite{cavitation pressure in water}. Cavities can be formed in fast flows of liquid when pressure drops to such a low level that vapor bubbles are formed \cite{textbook - Fundamentals of Fluid Mechanics}. They can also be produced by raising the local temperature of the fluid, as in laser-induced cavitation \cite{laser}. These cavities formed in local low-pressure zones collapse rapidly in the bulk liquid.

Formation of cavities in low pressure flows happens in natural as well as industrial settings. There is evidence that dolphins and fish experience cavitation when they swim fast and the snapping shrimp uses cavitation to stun its prey \cite{dolphins and fish, shrimp 1, shrimp 2}. In tracheids, trees' natural microfluidic pump system, water pressure can drop so low below atmospheric pressure that cavitation occurs, blocking water transport in that channel \cite{trees}. Near solid boundaries, collapsing cavities are known to form liquid jets that point toward the boundary \cite{cavitation near horizontal plane}. Due to these powerful and damaging liquid jets, in many cases cavitation in industrial settings is a negative occurrence -- it causes damage to ship propellers and pumps, and adds complications to flow in fuel injection systems \cite{propellers, pumps, fuel injection}. Cavitation has found use, however, in medical treatment, in ultrasonic cleaning, and for mixing or initiating chemical reactions in microfluidic systems \cite{medical treatment, kidney stones, ultrasonic cleaning 1, mixing 1, mixing 2, catalyzing chemical reactions}.

Cavitation near solid boundaries of different shapes has been studied, with the complexity of the boundary shape gradually increasing over time from single horizontal planes to parallel or perpendicular walls and curved boundaries \cite{parallel walls, cavitation near horizontal plane, cavitation near perpendicular walls, cavitation near wall and corners, cavitation near semi-enclosed parallel walls, cavitation near curved boundaries}. Various aspects of cavitation including bubble shape, jet strength, and jet direction have been studied \cite{scaling laws for jets, cavitation in corners, cavitation near boundary with attached bubble}. Jet direction has been studied near various complex boundaries \cite{cavitation near boundary with attached bubble, sawtooth boundary}, but exact analytic predictions of jet direction are lacking in all but the most symmetric boundary cases \cite{cavitation near curved boundaries}. In jet formation flows develop around the bubble surface with very high velocities in the direction of the liquid jet \cite{flow field, microfluidic triangle and square}. Moving away from the high degree of symmetry in previous studies, Tagawa and Peters (2018) predicted jet direction for bubbles collapsing near solid corners of angle $\pi / n$ (for natural number $n$). Their equation relating jet direction to position within the corner is an analytic solution based on modeling the flow field at the time of collapse initiation using potential flow analysis and the method of images \cite{cavitation in corners}. Here, we demonstrate that the potential flow method can be used to predict jet direction within several types of polygons, specifically, any rectangle, equilateral triangle, isosceles right triangle, and $30^{\circ}$-$60^{\circ}$-$90^{\circ}$ right triangle.

A bubble in a liquid region enclosed by solid walls is modeled as a sink in a potential flow. The boundary condition that fluid cannot flow through the walls is satisfied by introducing an infinitely tessellating pattern of image sinks. Experiments were conducted using laser-induced cavitation, and the models are found to agree with experimental data. Conceptually, this demonstrates that regardless of the complex dynamics of the bubble surface, asymmetry in the flow field in the initial instants of collapse is sufficient to predict the direction of the liquid jet of a collapsing bubble.

The method of images is a mathematical result that is widely applicable, including to our potential flow problem and any situation in which Laplace's equation is satisfied \cite{method of images}. Because of this, the analytic model presented has further significance because it demonstrates the applicability of method of images solutions to describe a physical situation with relatively complex, enclosed boundary conditions.

\section{Experiments}
\begin{figure}[htbp]
    \centering
     \includegraphics[width=84mm]{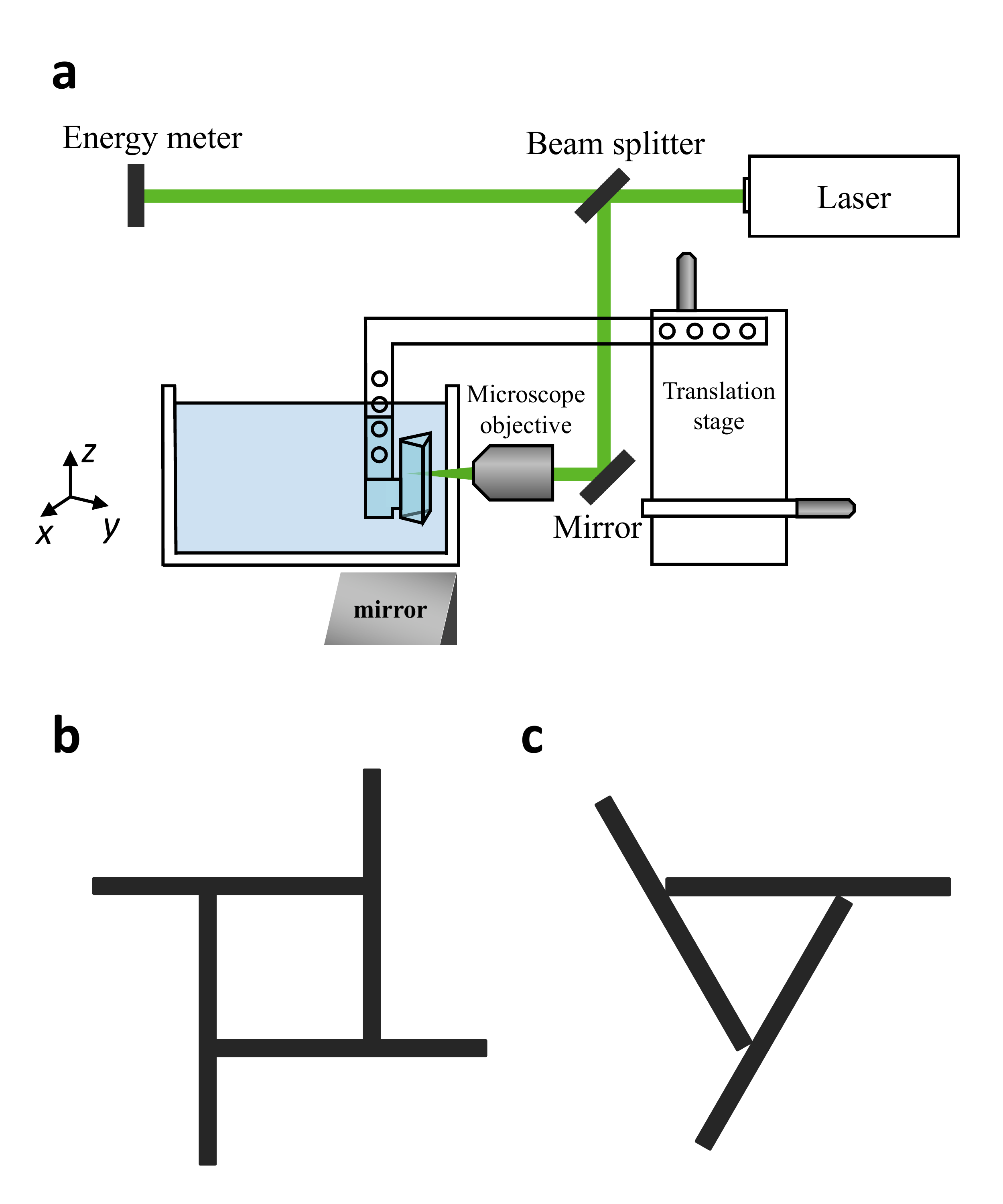}
    \caption{Experiment setup. (a) Bubbles are produced in a large water bath by laser-induced cavitation. The tank is transparent. The beam passes through a microscope objective, which focuses the laser to a point where the bubble is nucleated within a geometrically shaped microscope slide assembly. Images are taken of the $xy$-plane after being reflected by a mirror placed below the tank. (b) Square cross section. Inner side length: 15 mm. (c) Equilateral triangle cross section. Inner side length: 15 mm.}
    \label{fig:experiment_diagram}
\end{figure}

Bubbles were produced by a laser in a large ($145 \times 145 \times 145$ mm$^3$) acrylic tank filled with degassed water. Figure \ref{fig:experiment_diagram}a is a diagram of the experimental arrangement. Glass microscope slides of dimensions $25.7 \times 75.8 \times 1.4$ $\text{mm}^{3}$ were glued together to form geometric shapes as shown in Figure \ref{fig:experiment_diagram}b and \ref{fig:experiment_diagram}c. The slides formed a channel with the cross-section of a square or equilateral triangle of side length $l = 15 \text{mm}$, and were long enough to approximate an infinitely long tunnel with the desired cross-section. The channel is long enough (for channel length $L$ and bubble diameter $d$, we have roughly $L / d > 7$) to ignore any influence of the finite length, since, at large distances, changes in boundary condition have very weak effects on bubble jets \cite{cavitation near perpendicular walls}. In addition, bubbles were produced midway between the channel's two open ends, to reduce any effect of asymmetry in the $z$-direction.

The microscope slide assembly was mounted using an acrylic arm attached to a translation stage accurate within 5 $\mu \text{m}$ (Figure \ref{fig:experiment_diagram}a). The assembly was mounted vertically (open ends toward top and bottom) to allow remnants of bubbles from previous cavitation events to escape by buoyancy. The slide assembly was positioned roughly equidistant between the tank bottom and the water surface.

A Q-switched Nd:YAG laser (Bernoulli PIV; Litron) produces enough concentrated energy to vaporize a small amount of water and produce a vapor bubble \cite{laser}. Laser pulse duration was 6 ns and wavelength was 532 nm. Laser energy was halved using a 50:50 beam splitter, with half the beam directed toward an energy meter and the other half directed into the water tank. Laser energy was kept at about $5$ mJ per pulse. Using a $10 \times$ microscope objective (Nikon Plan Fluor; NA = 0.30; working distance in air = 16 mm; working distance in water = 21 mm; Airy disk radius = 1 $\mu \text{m}$), laser light was focused to a point within the microscope slide assembly after passing through the clear acrylic tank and the glass slide.

Images were taken of bubbles growing and collapsing within the geometric shapes using a high-speed camera (Photron FASTCAM SA-X2) with a 105 mm Nikon Micro-Nikkor lens, recording at 100 kHz with an image size of $364 \times 284$ pixels. The camera recorded a bottom view by using a mirror positioned below the water tank. A 550 nm longpass filter blocked laser light from entering the camera lens. The position and jet angle of the bubbles were measured from these images. We calibrated our images using a millimeters per pixel resolution measured by the difference in pixel position of the slide assembly between pairs of images where it was moved by a known millimeter displacement using the translation stage. The resolution was between 0.0295 mm per pixel (square experiment) and 0.0301 mm per pixel (triangle experiment). 

\begin{figure}[htbp]
    \centering
    \includegraphics[width=60mm]{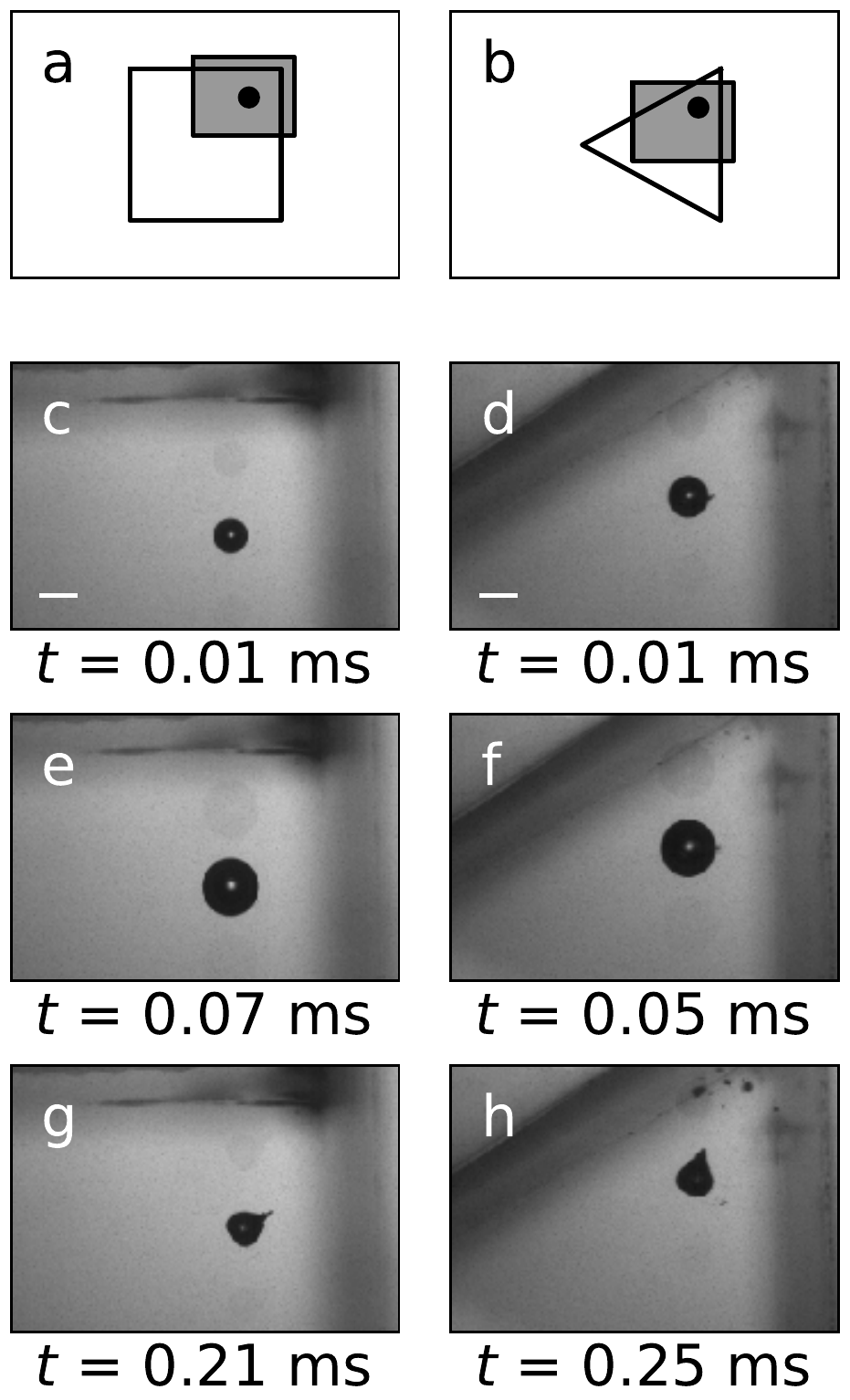}
    \caption{Bubble growth and collapse. Position of the camera frame in relation to the shapes is shown in (a, b). Bubbles are shown shortly after nucleation (c, d), at maximum area (e, f), and after a liquid jet has pierced the bubble (g, h). The diameters of the bubbles at maximum area are both roughly $d \approx 1.7 \text{mm}$. Scale bar is $1$ mm. Image brightness has been increased by 50\%.}
    \label{fig:movie_images}
\end{figure}

A typical set of bubble growth and collapse images from two movies is displayed in Figure \ref{fig:movie_images}. The bubble first expands spherically from the point of nucleation (Figure \ref{fig:movie_images}c and \ref{fig:movie_images}d), grows to its largest diameter (Figure \ref{fig:movie_images}e and \ref{fig:movie_images}f), and then collapses, at which point a liquid jet pierces through the bubble (Figure \ref{fig:movie_images}g and \ref{fig:movie_images}h).

The image analysis procedure involves measuring position and jet angle of each bubble. First a background image (without the bubble) is subtracted from each frame, and a binary threshold is applied to the image, defined so that the bubble appears in white and the rest of the image in black. A remaining hole on the bubble surface caused by the backlighting in the video is filled in. The bubble's area is calculated in each frame and converted from pixels to mm$^2$. The bubble position can then be measured. The bubble position is defined to be the bubble's centroid when it has reached its maximum size. The bubble is the most circular at this stage of expansion, so its center is best measured at this stage. The jet angle is determined by the direction of the bubble's displacement between its initial expansion and first rebound, which was found to be accurate compared to directly measured jet angles within 0.2 rad by Tagawa and Peters \cite{cavitation in corners}. The rebound position is determined by the bubble's centroid when it has reached its second maximum size. Although the bubble is not spherical at this point, it is symmetric around the jet axis (Figure \ref{fig:movie_images}g and \ref{fig:movie_images}h), so the centroid position still reflects the jet direction accurately.

Experiments were done along linear axes through the square and equilateral triangle, chosen based on behavior of the mathematical prediction of bubble jet direction in different regions of these shapes. Two sets of experiments were done for the square and four were done for the equilateral triangle.

\section{Model}
A collapsing bubble forms a jet when the velocity field around the bubble is not spherically symmetric, which occurs in the presence of solid boundaries \cite{microfluidic triangle and square, flow field}. Local regions of high fluid velocity appear to deform and indent a bubble's surface, leading to the eventual formation of a jet. Because of this, our model seeks to describe the fluid velocity at the bubble's surface during the initial part of the collapse.

We use potential flow theory to describe the fluid velocity field at the bubble surface. The flow field $\vec{U}$ of an inviscid, incompressible fluid such as water can be expressed as the gradient of a velocity potential $\phi$, such that $\vec{U}=\vec{\nabla}\phi$ \cite{textbook - Fundamentals of Fluid Mechanics} and such that $\phi$ satisfies Laplace's equation $\nabla^2\phi=0$.

We neglect viscosity and compressibility for the same reasons cited by Tagawa and Peters in similar experiments -- viscous effects are relevant only within thin boundary layers in the timescale of our experiments and the bubble surface speeds are minimal just after it has reached its maximum size \cite{cavitation in corners}.

\begin{figure}[htbp]
    \centering
    \includegraphics[width=70mm]{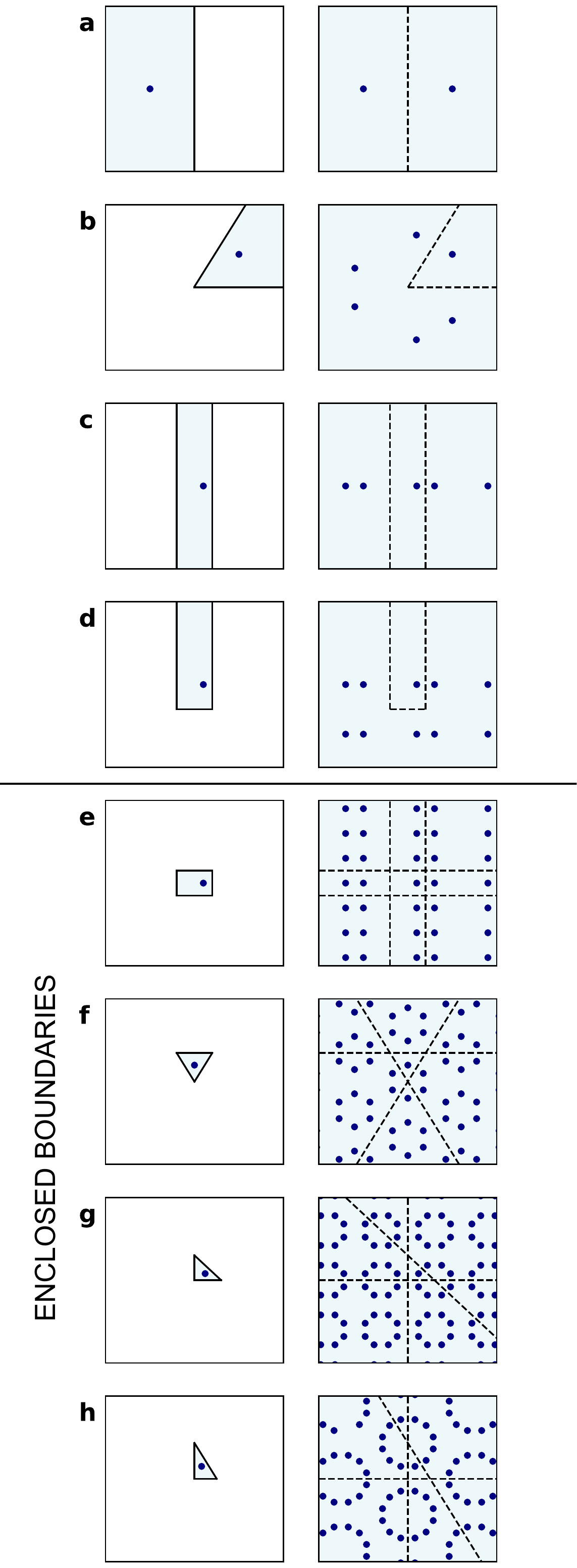}
    \caption{Modeling boundary conditions using method of image sinks. Image sinks are used to satisfy the solid boundary conditions for several different shapes: (a) flat wall, (b) $60^{\circ}$ corner, (c) parallel walls, (d) semi-enclosed parallel walls, (e) rectangle, (f) equilateral triangle, (g) isosceles right triangle, and (h) 30$^{\circ}$-60$^{\circ}$-90$^{\circ}$ right triangle. The patterns are symmetric around each of the boundaries (dashed lines), so there is no flow perpendicular to the boundary and the solid boundary condition is met.}
    \label{fig:image_sinks}
\end{figure}

We model the bubble as a three-dimensional point sink in the fluid and require the boundary condition that fluid cannot pass through any solid boundaries. A three-dimensional sink, as opposed to a two-dimensional sink is required because it correctly represents the radial fluid velocity produced by a spherically collapsing bubble \cite{cavitation in corners}. In Figure 3, the left column shows the different channel shapes we want to model. We can produce problems with identical boundary conditions using the method of images. By placing image sinks in symmetric, tessellating patterns, solid boundaries can be modeled.  Figure \ref{fig:image_sinks}a displays the simplest method of images solution in which a bubble and solid boundary are replaced by two sinks equidistant from an imaginary boundary plane (dashed line). Because of symmetry, fluid does not flow perpendicular to the plane of symmetry between the two sinks -- at any position on the plane, the sinks have equal and opposite pull in the direction perpendicular to the plane. In Figure \ref{fig:image_sinks}, solid boundaries of several shapes are modeled by patterns of image sinks: (a) a single wall, (b) a corner, (c) parallel walls, (d) semi-enclosed parallel walls, (e) a rectangle, (f) an equilateral triangle, (g) an isosceles right triangle, and (h) a $30^{\circ}$-$60^{\circ}$-$90^{\circ}$ right triangle. Each subsequent pattern follows the principle of the simple example in Figure \ref{fig:image_sinks}a: patterns are completely symmetric around each plane made by a solid boundary (dashed lines), so along these boundary planes there is no flow perpendicular to the plane. From Figure \ref{fig:image_sinks}a through Figures \ref{fig:image_sinks}c-e, for example, we see a progression of reflections across each successive boundary added so that the pattern remains symmetric until it is reflected across all four boundaries of a rectangle. Thus, the boundary condition is satisfied by these patterns of image sinks. 

Because of the uniqueness theorem, two physical setups that are identical within a region of interest and satisfy the same boundary conditions will produce the one and only unique fluid velocity field $\vec{U}$ within that region (for flow which is singly connected) \cite{method of images}. For example, in Figure \ref{fig:image_sinks}f, the region of interest -- which contains a point sink within an equilateral triangle fluid region -- is identical in both diagrams, and both physical setups satisfy a boundary condition that no fluid flows across the triangular boundary. For the specified solid boundaries in the left column, tessellating image sink patterns in the right column produce the exact same descriptions of the fluid velocity field that would occur, except they do so using point sinks, which are easier to describe analytically than a spherical bubble near complex solid boundaries. Although we are not the first to identify these method of images solutions \cite{periodic image cloud, cavitation near wall and corners}, here we show that we can apply this to the study of cavitation in polygonal channels.

In the remainder of this section, we will determine the equations representing sink positions in each pattern (A), calculate the velocity field induced by the sinks (B), and find the jet angle as a function of bubble position (C). Image sink positions for the four polygons are given by the equations below, although different formulations may be devised for the same patterns.

\subsection{Sink position}

\subsubsection{Rectangle}
In our model of a collapsing bubble within a rectangle having side lengths $l_x$ and $l_y$, sink positions are given by
\begin{align}
    \vec{s}_{ij} &= x_{s_{ij}} \hat{x} + y_{s_{ij}} \hat{y} \notag \\
    &= [(-1)^i \cdot x_b + i \cdot l_x] \hat{x} + [(-1)^j \cdot y_b + j \cdot l_y] \hat{y} \label{eq:sink_positions_sq}
\end{align}
for $-\infty < i < \infty$ and $-\infty < j < \infty$. For a square, we use equal side lengths, $l_x = l_y = l$. The position given by $i = j = 0$ is the bubble position itself, $\vec{s}_{00} = x_b \hat{x} + y_b \hat{y}$, while the rest of the positions are image sinks introduced to fulfill the boundary conditions at the rectangle boundaries.

\subsubsection{Equilateral triangle}
In our model of a collapsing bubble within an equilateral triangle of side length $l$, sink positions are given by
\begin{align}
    \vec{s}_{ijk} &= x_{s_{ijk}} \hat{x} + y_{s_{ijk}} \hat{y} \notag \\
    &= [a_{ijk} \cos (\frac{k \pi}{3}) - b_{ijk} \sin(\frac{k \pi}{3}) + c_{ijk} \cos(\pi + \frac{k \pi}{3})] \hat{x} 
    \notag \\
    &+ [a_{ijk} \sin(\frac{k \pi}{3}) + b_{ijk} \cos(\frac{k \pi}{3}) + c_{ijk} \sin(\pi + \frac{k \pi}{3})] \hat{y},
    \label{eq:sink_positions_tri_y}
\end{align}
where
\begin{align}
    a_{ijk} &= x_b + i \cdot \frac{3l}{2} - k^2 x_b
    \label{eq:velocity_field_tri_p} \\
    b_{ijk} &= (-1)^j \cdot (y_b - \frac{h}{2}) + j \cdot h + \frac{-h + (-1)^i h}{2} + \frac{h}{2} 
    \label{eq:velocity_field_tri_q} \\
    c_{ijk} &= k^2 x_b
    \label{eq:velocity_field_tri_r}
\end{align}
for $k \in \{-1, 0, 1 \}$, $-\infty < i < \infty$, and $-\infty < j < \infty$. The height $h$ is the triangle height $h = \sqrt{l^2 - (l/2)^2}$. The three values of $k$ represent rotations, each separated by $60^{\circ}$, of a base pattern used to construct a hexagonal pattern of image sinks.
The position given by $i = j = k = 0$ is the bubble position itself, $\vec{s}_{000} = x_b \hat{x} + y_b \hat{y}$, while the rest of the positions are image sinks introduced to fulfill the boundary conditions at the triangle boundaries.

\subsubsection{Isosceles right triangle}
In our model of a collapsing bubble within an isosceles right triangle having side length $l$, sink positions are given by
\begin{align}
    \vec{s}_{ijk} &= x_{s_{ijk}} \hat{x} + y_{s_{ijk}} \hat{y} \notag \\
    &= \left[d_{ij} \cos(\frac{k \pi}{2}) - f_{ij} \sin(\frac{k \pi}{2}) + l \right] \hat{x} \notag \\
    &+ \left[d_{ij} \sin(\frac{k \pi}{2}) + f_{ij} \cos(\frac{k \pi}{2}) \right] \hat{y} \label{eq:sink_positions_isosceles_right}
\end{align}
where
\begin{align}
    d_{ij} &= (-1)^i \cdot \left(x_b - \frac{l}{2} \right) + i \cdot l - \frac{l}{2}
    \label{eq:velocity_field_isosceles_right_x} \\
    f_{ij} &= (-1)^j \cdot \left(y_b - \frac{l}{2} \right) + j \cdot l + \frac{l}{2}
    \label{eq:velocity_field_isosceles_right_y}
\end{align}
for $k \in \{0, 1 \}$, $-\infty < i < \infty$, and $-\infty < j < \infty$. The two values of $k$ represent rotations, separated by $90^{\circ}$, of a base pattern used to construct the full pattern of image sinks. As with the equilateral triangle, the position given by $i = j = k = 0$ is the bubble position itself, $\vec{s}_{000} = x_b \hat{x} + y_b \hat{y}$.

\subsubsection{30$^{\circ}$-60$^{\circ}$-90$^{\circ}$ triangle}

In our model of a collapsing bubble within a 30$^{\circ}$-60$^{\circ}$-90$^{\circ}$ triangle having hypotenuse of length $l$ (short side length $\frac{l}{2}$ and long side length $h = \sqrt{l^2 - (l/2)^2}$), sink positions are given by
\begin{align}
    \vec{s}_{ijk} &= x_{s_{ijk}} \hat{x} + y_{s_{ijk}} \hat{y} \notag \\
    &= \left[m_{ij} \cos(\frac{k \pi}{3}) - n_{ij} \sin(\frac{k \pi}{3}) \right] \hat{x} \notag \\
    &+ \left[m_{ij} \sin(\frac{k \pi}{3}) + n_{ij} \cos(\frac{k \pi}{3}) + h \right] \hat{y}
\end{align}
where
\begin{align}
    m_{ij} = (-1)^i \cdot (x_b - \frac{3l}{4}) + i \cdot \frac{3l}{2} + (-1)^q \cdot \frac{3l}{4} \\ 
    n_{ij} = (-1)^j \cdot (y_b - \frac{h}{2}) + j \cdot h - (-1)^q \cdot \frac{h}{2}
\end{align}
for $k \in \{-1, 0, 1\}$, $q \in \{0, 1\}$, $-\infty < i < \infty$, and $-\infty < j < \infty$. The three values of $k$ represent rotations, separated by $60^{\circ}$, of a base pattern used to construct the full pattern of image sinks. Both positive and negative values of $\frac{3l}{4}$ and $\frac{h}{2}$ are required to produce an offset section that completes the pattern, hence the $(-1)^q$ coefficient. As with the previous two cases, the position given by $i = j = k = 0$ is the bubble position itself, $\vec{s}_{000} = x_b \hat{x} + y_b \hat{y}$.

\subsection{Velocity field}

The fluid velocity field $\vec{u}_s (x, y)$ induced by one 3D point sink at a position $\vec{s} = (x_s, y_s)$ is
\begin{equation}
    \vec{u}_s (x,y) = -\frac{Q}{4 \pi R^3}[(x - x_s) \hat{x} + (y - y_s) \hat{y}]
\end{equation}
where $Q$ is the sink strength in terms of volume per unit time and $R = \sqrt{(x - x_s)^2 + (y - y_s)^2}$ is distance from the sink position.

For a set of image sink positions $\{s_{ij}\}_{i, j \in \mathbb{Z}}$, the velocity field induced by all of these sinks is the sum of the velocity fields induced by each individual sink. This velocity field for the square is given by
\begin{align}
    \vec{u} (x, y) 
    &= \sum_{\substack{i, j = -\infty \\ (i, j) \neq (0, 0)}}^{\infty} \vec{u}_{s_{ij}} \label{eq:velocity_field_sq}
\end{align}
for the isosceles right triangle is given by
\begin{align}
    \vec{u} (x, y) 
    &= \sum_{\substack{i, j = -\infty \\ k \in \{0, 1 \} \\ (i, j, k) \neq (0, 0, 0)}}^{\infty} \vec{u}_{s_{ijk}} \label{eq:velocity_field_isc}
\end{align}
and for the equilateral triangle and $30^{\circ}$-$60^{\circ}$-$90^{\circ}$ triangle is given by
\begin{align}
    \vec{u} (x, y) 
    &= \sum_{\substack{i, j = -\infty \\ k \in \{-1, 0, 1 \} \\ (i, j, k) \neq (0, 0, 0)}}^{\infty} \vec{u}_{s_{ijk}} \label{eq:velocity_field_eq_306090}
\end{align}
where sink positions $\vec{s}_{ij}$ for a square, or $\vec{s}_{ijk}$ for an equilateral triangle, isosceles right triangle, or $30^{\circ}$-$60^{\circ}$-$90^{\circ}$ triangle are as given in Equations $(1$-$11)$. We have excluded the velocity field produced by the bubble itself by the condition $(i, j) \neq (0, 0)$ or the condition $(i, j, k) \neq (0, 0, 0)$, as applicable. This is because the flow field induced by the bubble itself is radially symmetric around the bubble position, so it is the resultant velocity from the \textit{image} sinks, not the bubble, that deforms its surface.

\subsection{Jet direction}

We assume that the resultant velocity induced at the bubble position by all image sinks gives the direction of the jet, meaning the jet direction $\theta_j$ is given by
\begin{equation}
    \tan{\theta_j} = \frac{u_y}{u_x} \label{eq:jet_angle}
\end{equation}
where $u_x$ and $u_y$ are the velocity components at the bubble position; that is, they are the $x$ and $y$ components of $\vec{u}(x, y)|_{(x, y) = (x_b, y_b)}$ as given by Equations (13-15) \cite{Note1}.

\begin{figure}[htbp]
    \centering
    \includegraphics[width=84mm]{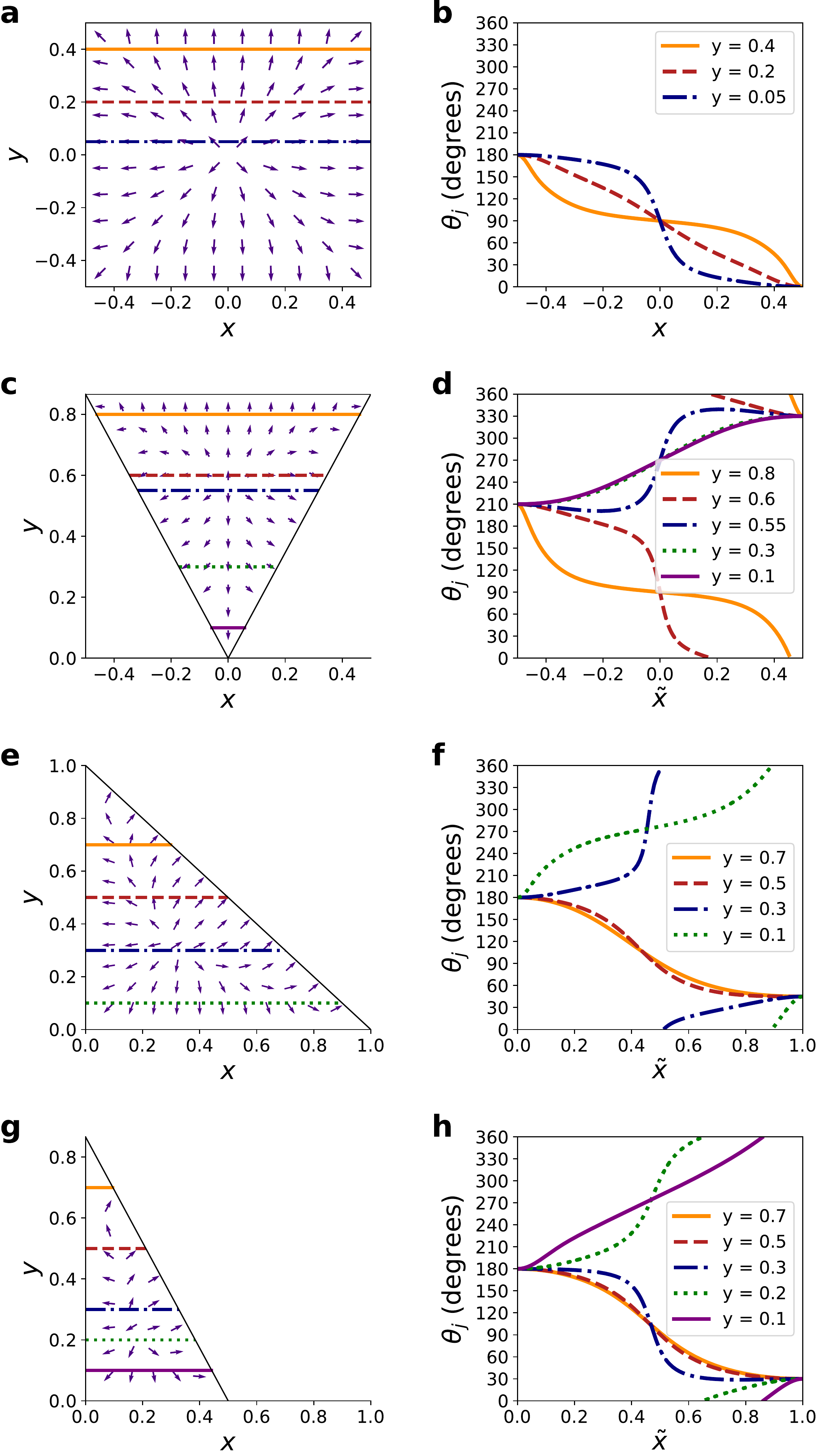}
    \caption{Model predictions of jet direction. Polygons are (a, b) a square, (c, d) an equilateral triangle, (e, f) an isosceles right triangle, and (g, h) a $30^{\circ}$-$60^{\circ}$-$90^{\circ}$ triangle. Each vector within the polygons represents the jet direction produced by a bubble nucleated at that position (a, c, e, g). Plotting predictions along continuous lines within the shapes produces prediction curves (b, d, f, h), which can be tested experimentally. Side length of each shape is chosen to be $1$, except for (g), which is plotted so it is half the equilateral triangle's size (c). Prediction curves for the triangles (d, f, h) are plotted against a position variable $\widetilde{x}$ normalized by length of the line for ease of comparison.}
    \label{fig:model_plots}
\end{figure}

The jet direction $\theta_j$ for a bubble collapse at any point in the fluid can be calculated using Equation \eqref{eq:jet_angle}. Calculations of jet direction for bubble collapse at different points within the four polygons are displayed by vector fields in Figure \ref{fig:model_plots}a, \ref{fig:model_plots}c, \ref{fig:model_plots}e, and \ref{fig:model_plots}g. The vector fields in Figure \ref{fig:model_plots} are \textit{not} fluid velocity fields, but display the jet direction for bubbles nucleated at different points within the shape. Thus, Figure \ref{fig:model_plots} represents the complete model of analytic predictions of jet direction in bubble collapse within a square, an equilateral triangle, an isosceles right triangle, and a $30^{\circ}$-$60^{\circ}$-$90^{\circ}$ triangle. From this analytic model, we can make general observations about bubble collapse within these four polygonal shapes.

\subsection{3D sink model behavior}
The model confirms our expectation that bubble jets point perpendicular to solid boundaries when close to the boundaries. In other words, bubble jets generally point toward the closest wall or walls. We also observe from the jet direction arrows in Figure \ref{fig:model_plots}a that along axes of symmetry of the square, the jet points along the symmetry axis. This behavior can also be observed within the equilateral triangle (Figure \ref{fig:model_plots}c) and isosceles right triangle (Figure \ref{fig:model_plots}g). The 30$^{\circ}$-60$^{\circ}$-90$^{\circ}$ triangle does not have a symmetry axis. This bubble behavior along symmetry axes is similar to that observed along the symmetry axis of all corners of angle $\alpha = \frac{\pi}{n}$ \cite{cavitation in corners}.

Model curves can be plotted along straight lines within the shapes. Representative curves for all shapes are plotted in Figure \ref{fig:model_plots}b, \ref{fig:model_plots}d, \ref{fig:model_plots}f, and \ref{fig:model_plots}h. Calculating  predicted jet directions at different positions along straight lines within shapes is one way to see the change in the respective influence of the solid walls at varied positions. At the left and right edges of the graphs in Figure \ref{fig:model_plots}b, \ref{fig:model_plots}d, \ref{fig:model_plots}f, and \ref{fig:model_plots}h, for example, all predictions converge to the same angle such that jets are perpendicular to the leftmost or rightmost wall, demonstrating that the influence of a single wall dominates at positions very close to the wall. Similarly, along the top edge of the square (Figure \ref{fig:model_plots}a and \ref{fig:model_plots}b, $y = 0.4$) and equilateral triangle (Figure \ref{fig:model_plots}c and \ref{fig:model_plots}d, $y = 0.8$), the jet angle hovers around $\theta_j = 90^{\circ}$, except closer to the left and right edges, where the jet is influenced by the other boundaries. Closer to the center of the shapes (square -- $y = 0.05$, triangle -- $y = 0.55$ or $y = 0.6$), a sharp transition in jet angle occurs when the dominant influence switches from the left boundary to the right one.

\begin{figure}[htbp]
    \centering
    \includegraphics[width=84mm]{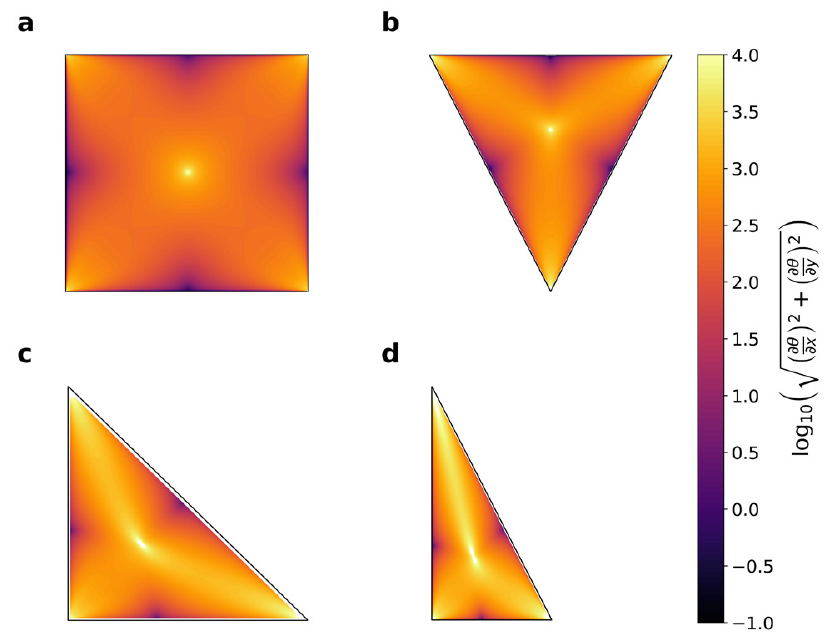}
    \caption{Volatility of jet direction angle. Within the (a) square, (b) equilateral triangle, (c) isosceles right triangle, and (d) $30^{\circ}$-$60^{\circ}$-$90^{\circ}$ triangle, a bubble's jet angle changes faster over small distances in the center and corners of the shape. In these regions, a bubble's jet direction is very sensitive to its position. Here we quantify volatility as the magnitude of the gradient of the jet angle, plotted as the logarithm.}
    \label{fig:volatility}
\end{figure}

Additionally, the model predicts jet direction to be highly sensitive to changes in position in the middle of these shapes and, to a lesser degree, at the corners. These are areas where a large change in jet angle may occur with only a small change in position. We call this sensitivity to changes in position the `volatility' in jet direction, which varies by location. Volatility is predicted by the model to be highest near the center and corners of the shapes, as depicted in Figure \ref{fig:volatility}. Here we quantify volatility as the magnitude of the gradient of the jet angle (plotted as the logarithm).

These observations may be used to apply jets to specific directions within microfluidic channels of geometric cross sections, at least when the channel is large enough that the approximation of the bubble as a point sink is reasonable.

\section{Comparison with Experiment}
\begin{figure}[htbp]
    \centering
    \includegraphics[width=84mm]{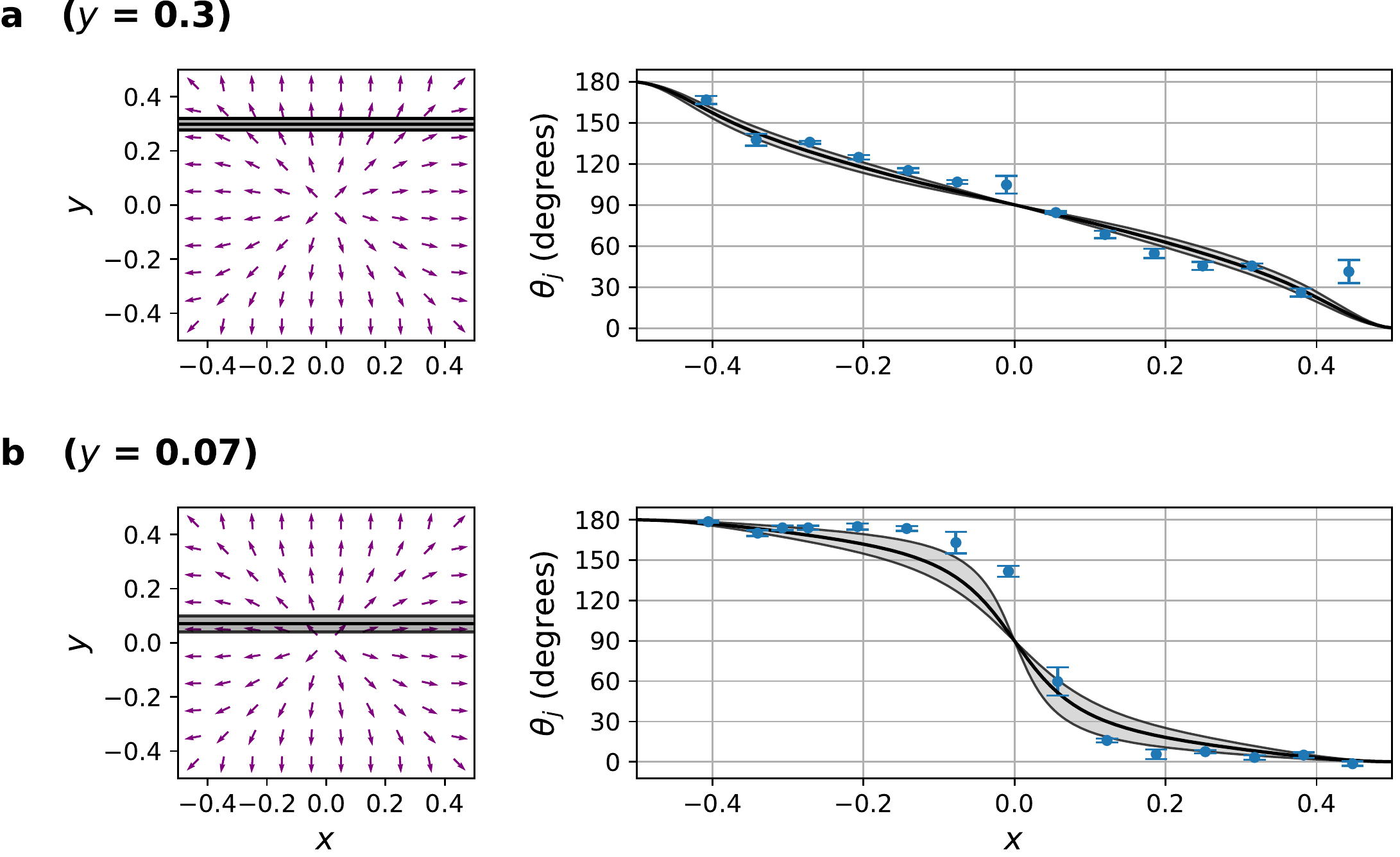}
    \caption{Square channel results. The jet angle of bubbles collapsing at several points along the axes (a) $y = 0.3$ and (b) $y = 0.07$ within a square channel was measured. Data points are the mean of several trials. Error bars are standard error of the mean. The model is plotted in a solid black line. Because bubble nucleation did not occur exactly on the intended axis, this is taken into account in how the model is displayed. Shaded regions represent model predictions resulting from moving two standard deviations above and below the mean bubble $y$ position.}
    \label{fig:square_results}
\end{figure}

\begin{figure}[htbp]
    \centering
    \includegraphics[width=84mm]{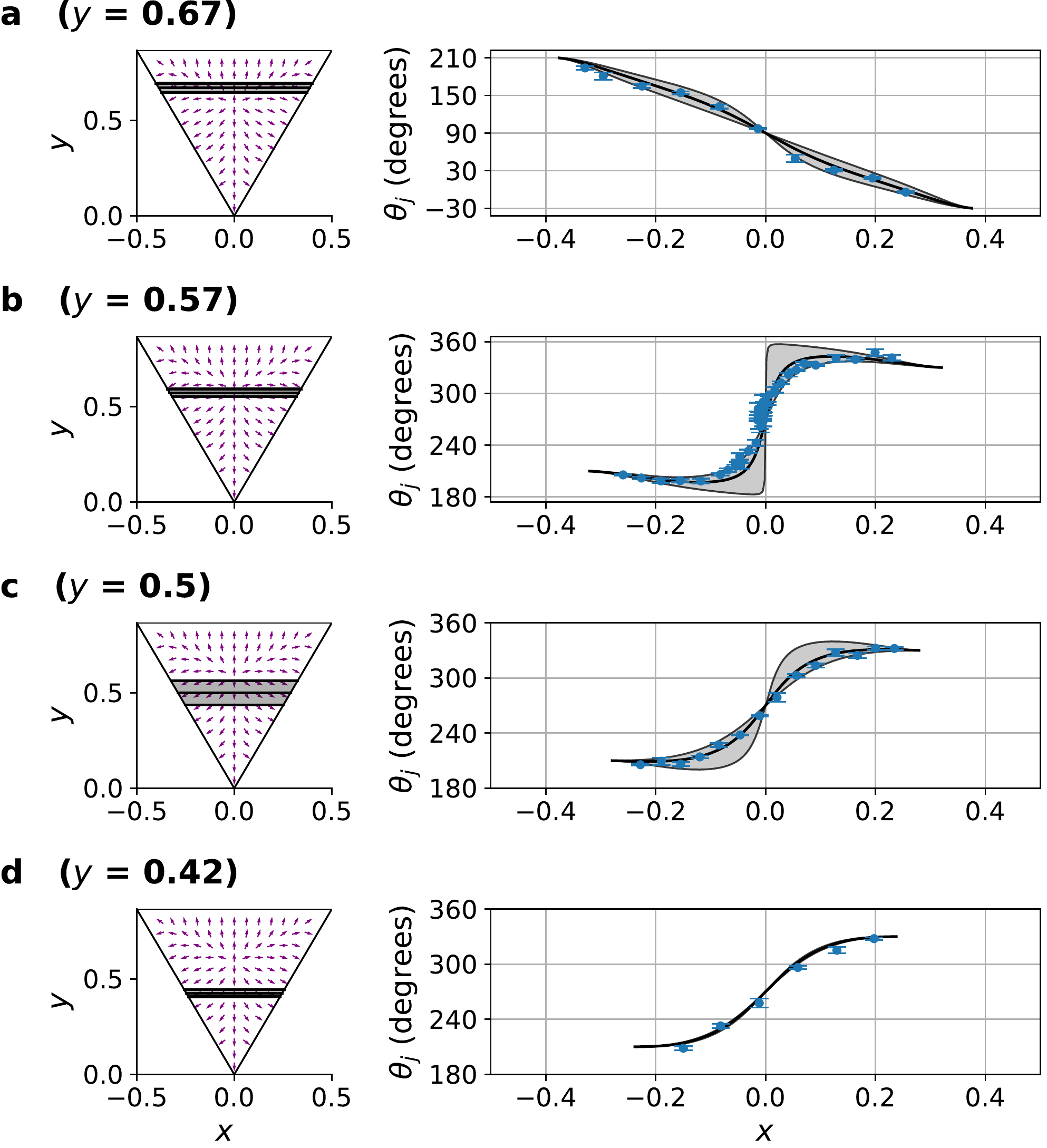}
    \caption{Equilateral triangular channel results. The jet angle of bubbles collapsing at several points along the axes (a) $y = 0.67$, (b) $y = 0.57$, (c) $y = 0.5$, and (d) $y = 0.42$ within an equilateral triangular channel was measured. Data points are the mean of several trials. Error bars are standard error of the mean. The model is plotted in a solid black line. Because bubble nucleation did not occur exactly on the intended axis, this is taken into account in how the model is displayed. Shaded regions represent model predictions resulting from moving two standard deviations above and below the mean bubble $y$ position.}
    \label{fig:triangle_results}
\end{figure}

Results from experiments within the square channel and the equilateral triangle channel are plotted in Figures \ref{fig:square_results} and \ref{fig:triangle_results}. For each of the six experiments, we measured bubble jet direction at a series of positions along a single line within the shape.  The positions $x$ and $y$ are non-dimensional lengths, meaning bubble positions (blue data points) are normalized by the side length of the shapes. A large number of trials were conducted at each position \cite{Note2}. Data points display the mean jet angle over these trials in the same position. Error bars are standard error of the mean. The model is plotted in a solid black line and is found to be in agreement with the data. Because bubbles are nucleated at positions that deviate slightly from the intended axis, the spread in position must be considered in how the model is displayed. Shaded regions in Figure \ref{fig:square_results} and Figure \ref{fig:triangle_results} represent the model prediction that results from considering the distribution of bubble $y$-positions and moving two standard deviations above and below the mean bubble $y$ position. In each experiment, the shaded region should thus represent the analytic prediction for 98\% of bubbles produced. Deviations in $y$ position tend to be small, $\pm 2$-$3 \%$ of the shape dimension, except in the case of Figure \ref{fig:triangle_results}c, where the deviation in $y$ position is $\pm 6 \%$ of the shape dimension.

Across all experiments bubbles had an average diameter of 1.4 mm. Bubbles within the equilateral triangle had a slightly smaller average diameter, 1.36 mm, compared to the average diameter of bubbles within the square, 1.48 mm, but the difference is within a standard deviation of ~0.3 mm. A typical errorbar length is 2-4 degrees, with the exception of Figure \ref{fig:triangle_results}c, in which errorbars have an average length of 9 degrees (despite having a similar number of trials to the other experiments). This suggests that, as the model predicts (Figure \ref{fig:model_plots}c), jet angle is much more volatile closer to the middle of the equilateral triangle shape. We also note that near the center of the square shape (Figure \ref{fig:square_results}b, $y = 0.07$), data has a significant deviation from the model around $x = 0$.

A visual inspection of the data compared to the model reveals a fair to good match. Quantitatively, the root-mean-square-deviations between the experimental points and the model are 11 degrees and 16 degrees in Figures \ref{fig:square_results}a and \ref{fig:square_results}b, respectively, and 9, 21, 5, and 5 degrees in Figures \ref{fig:triangle_results}a, \ref{fig:triangle_results}b, \ref{fig:triangle_results}c, \ref{fig:triangle_results}d, respectively.

\section{Discussion \& Conclusion}
We measured the direction of liquid jets produced by collapsing vapor bubbles in fluid enclosed by channels with a square or equilateral triangular cross-section. Experiments demonstrated that the theoretical model based on potential flow analysis agrees well with data. In total we have developed analytic predictions of jet direction for bubble collapse within any rectangle (including squares as a special case), equilateral triangle, isosceles right triangle, and $30^{\circ}$-$60^{\circ}$-$90^{\circ}$ triangle. Previous research modeled jet direction of collapsing cavities within corners by satisfying solid boundary conditions using the method of images. Our solutions extend this method using infinitely repeating symmetric patterns, and allow for the prediction of jet direction within enclosed channels with polygonal cross-section. The model is a powerful tool for analytically predicting jet direction within certain polygonal channels.

Our results reinforce Tagawa and Peters' claim that the velocity field around a bubble in the very beginning of its collapse is responsible for its later migration and jet direction \cite{cavitation in corners}. In addition, we introduce an analytic solution for enclosed cavitation, which has previously been studied only by experiment \cite{microfluidic triangle and square, bubble in a tube} or numerical analysis \cite{numerical calculation for a completely enclosed bubble}.

It is a key physical insight that the model does not describe the motion of the entire bubble surface throughout the collapse. Despite the complex shapes a bubble takes during collapse, describing the direction of its jet does not require a detailed model of the bubble surface boundary throughout the time of the collapse process. The bubble’s jet direction can be accurately predicted simply by observing the asymmetric flow field at its surface at the beginning of its collapse, which is the only thing our model takes into account.

Bubbles in our experiments had a typical diameter of $0.1l$, where $l$ is the side length of the shape. No atypical deformations are observed for bubbles this size besides the usual jetting expected near a solid boundary. It is expected that with larger bubbles interesting deformations will occur. For example, our experiments resemble the cavitation experiments in microfluidic systems performed by Zwaan et al. \cite{microfluidic triangle and square}. In their case, the bubble was of comparable size to the enclosing solid boundary shape and the bubble's collapse was considered a two-dimensional process. Deformations as well as multiple jetting was observed. Their experimental data was compared to a numerical model. Here, the bubble's collapse is three-dimensional and an analytic solution predicting jet direction is presented.

A goal of this research was to use a simple method to predict the direction of bubble jets in increasingly complex physical situations closer to those that may be encountered in human-made devices or in nature. In the realm of human-made devices, this research improves understanding of a bubble's collapse behavior within enclosed channels, and this is, for example, applicable to microfluidic channels with a polygonal cross-section. The results are a starting point for studying how cavitation may be used more effectively in mixing or cleaning in channels of different designs. Since the method of images is less computationally expensive than numerical solutions \cite{periodic image cloud}, it is a practical tool for understanding these applications.

It is of theoretical benefit to find the most complex situations for which the relatively simple method of images can work. The method of images provides a better theoretical understanding of the influence of the geometry of the solid walls on collapsing bubbles' jet formation. It appears that with the patterns presented in Figure \ref{fig:image_sinks} we have reached or are close to the limit of the method of images solutions under the conditions that the pattern (i) uses some (possibly infinite) number of point sinks that (ii) have equal strength and (iii) are arranged in a two-dimensional planar pattern. The authors suspect that the rectangle, equilateral triangle, isosceles right triangle, and $30^{\circ}$-$60^{\circ}$-$90^{\circ}$ triangle are the only enclosed polygons that can be modeled this way. This is because it seems no more shapes satisfy conditions required for this method to be applicable. First, we know that we cannot model polygons with more than four sides -- for example, regular pentagons, hexagons, heptagons, etc. -- because they have obtuse corner angles. For obtuse corner angles (angle greater than $\frac{\pi}{2}$), proper reflection of image sinks across each boundary eventually means that image sinks are placed within the region of interest rather than the boundary region, which will not produce the solution of a single bubble within the fluid region enclosed by the shape. Second, each solid boundary must have a mirror reflection of the image sink pattern on each side of the boundary plane. In practice, we have found that this means the boundary shape must be able to fully tessellate the plane. Only a small set of \textit{regular} shapes, the equilateral triangle, square, and hexagon, tessellate the plane. 
(We have not modeled the hexagon because of its obtuse corner angles.) Irregular polygons were also explored. However, they only satisfy the second condition (to tessellate the plane) in a few cases, which are related nicely to regular polygons: the rectangle, isosceles right triangle, and $30^{\circ}$-$60^{\circ}$-$90^{\circ}$ triangle (see Fig. \ref{fig:image_sinks}). Considering these conditions, it appears that we have identified the four special enclosed polygonal channel shapes for which the method of images works to predict jet directions for collapsing bubbles.

From a theoretical standpoint, these results encourage finishing the catalog of method of images solutions. It appears that we may have reached the limit of two-dimensional shapes that can be modeled by patterns of equal strength image sinks. However, there remain extensions to three dimensional shapes such as tetrahedra and dodecahedra, which can tessellate in three-dimensional space, as well as simpler extensions of the present results such as modeling cavitation within half-enclosed and fully-enclosed rectangular and triangular prisms.

From a computational standpoint, there are many avenues by which these results can be expanded. While the method of images provides geometrical limitations if we desire analytical solutions, there are no such geometrical limitations for numerical methods. For bubble collapse within any shape, the flow produced by a point sink in a fluid domain with the appropriate solid boundary conditions could be solved numerically and jet direction could be predicted by finding the resultant fluid velocity at the bubble position.

In summary:

(1) We have demonstrated a method to predict jet direction of collapsing bubbles within polygonal channels.

(2) Jet direction is more volatile in the middle and corners of the channel.

(3) Jets near the flat channel walls away from the corners will be perpendicular to the walls.

(4) The authors suspect that the four polygons presented are the only enclosed channel cross-sections that can be modeled using tessellating image sink patterns if the sinks are of equal strength.

\begin{acknowledgments}
We thank Elijah Andrews and Yoshiyuki Tagawa for insightful discussions. We acknowledge financial support from the EPSRC under Grant No. EP/P012981/1. LM acknowledges financial support from the Jeff Metcalf Global Fellowship Grant.
\end{acknowledgments}

\end{document}